\documentclass[showpacs,twocolumn,nofootinbib]{revtex4}%
\usepackage{amssymb}
\usepackage{amsmath}
\usepackage{amsfonts}
\usepackage{graphicx}%
\begin{document}
\title{Feasibility of Large Free-standing Liquid Films in Space}
\author{Rui Zheng }
\email{ruizheng@uchicago.edu}
\author{Thomas A. Witten}
\email{t-witten@uchicago.edu}
\affiliation{James Franck Institute and Department of Physics, The University of Chicago,
929 E. 57th Street, Chicago, Illinois 60637, USA}
\date{\today}

\begin{abstract}
We consider the feasibility of large-scale free-standing thin liquid film
experiment in the space environment as a new realization to study
two-dimensional hydrodynamics. We identify material and environmental criteria
necessary to avoid freezing, evaporation, chemical degradation, and
spontaneous collapse of the film. These criteria pose no obstacles to
achieving films of kilometer scale and lifetime of many months, with
attainable Reynolds number up to $10^{7}$. However, impacts from meteoroids
pose a serious threat to the film, and require substantial shielding or
unproven self-healing properties in the film. Current theoretical and
experimental studies of two-dimensional turbulence are briefly reviewed. We
also describe a specific candidate liquid for the film.

\end{abstract}
\maketitle

\section{Introduction}

Macroscopic thin liquid films are prevalent in nature and technology. To
understand their physical properties is theoretically important and
experimentally instructive \cite{1,103,104,102}. Thin liquid films are useful
tools for industrial applications and scientific explorations \cite{01}. In
particular they are conventional experimental realizations to study
two-dimensional hydrodynamics and turbulence \cite{200}.

Two-dimensional hydrodynamics plays a special role in space science
\cite{201}. Two-dimensional flows with high Reynolds numbers may organize
spontaneously into large-scale coherent patterns \cite{202, 203} that are
common features of geophysical and astrophysical flows. Some phenomena, such
as surface-tension-driven flows and thermocapillary flows, have important
applications in space-related techniques \cite{204}.

The space environment, on the other hand, provides some unique conditions for
experimental physics. In particular, the micro gravity and vacuum in space are
ideal environments for large-scale experiments that are not feasible in the
normal laboratory environment.

In this paper we propose a novel realization of a large-scale free-standing
liquid film in the space environment to study two-dimensional hydrodynamics.
Compared to conventional experimental realizations, the film-in-space
experiment is appealing for the following reasons. First, among the factors
causing the rupture of macroscopic thin films under normal laboratory
conditions, the primary factor is drainage due to gravity. We expect the
micro-gravity environment will extend the film lifetime against the
drainage-induced rupture. Second, the space environment makes it possible to
study large-scale flows in the film. In particular the environment permits
flows of higher Reynolds number and hence an improved study of two-dimensional
turbulence. Lastly, in current experiments based on conventional soap films
\cite{3} and other realizations \cite{200}, the underlying dynamics inevitably
couples with the adjacent matter such as gaseous and solid boundary layers.
These couplings create technical difficulties and cloud interpretation of the
experiments. The space films can readily help overcome these obstacles and
improve our understanding of two-dimensional hydrodynamics, with its
distinctive form of nonlinearity and turbulence.

This paper assesses the potential for creating, maintaining, and manipulating
large-scale free-standing liquid films in the space environment. We shall base
our study on conventional stability theory of soap films \cite{7}. The design
of the space film requires an unusual combination of fluid properties, such as
extremely low vapor pressure and appropriate viscosity and surface tension, in
order to maintain the film in a stable liquid state. After initial assessment
of these properties, we consider explicitly the commercial liquid \textit{Dow
Corning }705 (DC 705) diffusion pump oil \cite{29} as a possible candidate
film liquid. DC 705 oil (penta-phenyl-tri-methyl-tri-siloxane) is a colorless
to straw-colored, single component fluid designed for ultrahigh vacuum
applications. Conventional pump oils such as this give extremely low
evaporation rates with minimal increase in viscosity. We include its detailed
physical and chemical properties in Appendix A.

For definiteness, we assume the space film to be a circular film with diameter
$L=1%
%TCIMACRO{\unit{km}}%
%BeginExpansion
\operatorname{km}%
%EndExpansion
$ and thickness $h=1%
%TCIMACRO{\unit{\U{3bc}m}}%
%BeginExpansion
\operatorname{\mu m}%
%EndExpansion
$, maintained at a temperature $T_{0}=298%
%TCIMACRO{\unit{K}}%
%BeginExpansion
\operatorname{K}%
%EndExpansion
$ over a lifetime of a year. Accordingly the total mass of the base liquid is
of the order of $10^{3}%
%TCIMACRO{\unit{kg}}%
%BeginExpansion
\operatorname{kg}%
%EndExpansion
$. Based on these assumed parameters and the candidate liquid properties, we
can study the feasibility of the film under different external influences
expected from the space environment and suggest potential difficulties and
possible solutions. Moreover using these parameters as a base case and known
scaling laws, one may readily estimate effects of changing these parameters.

We also consider as an alternative the \textquotedblleft black
film\textquotedblright\ \ \cite{4}, which is a stable film configuration with
thickness of the order of a few hundred angstroms. The black film\ therefore
deviates from the standard stable thin-liquid-film model and our assumed
physical conditions, and requires alternative feasibility studies that yield
different estimates. We shall briefly discuss the advantages and shortcomings
of the black film.

In the next section we prescribe the thermodynamic properties required to
maintain the thin film in a liquid state with little evaporation. In Section
III we consider the stability of the film against spontaneous rupture and
discuss the requisite properties of the compatible surfactants that ensure the
stability. We then analyze the possible damages to the film from external
influences in the space environment in Section IV. In Section V we review the
background for\ two-dimensional hydrodynamics and turbulence, and estimate
parameters such as the Reynolds number achievable for the film. We also
compare the space film with current experimental realizations in Section VI.
For completeness, we consider in Sections VII and VIII various forces, both
external and internal, and time scales relevant for the film design. In the
Discussion section we suggest some alternatives to the standard film model and
possible generalizations to other experiments in the space environment. We
present our conclusion in the final section.

\section{Requisite thermodynamic properties to maintain the liquid state in
space}

\subsection{Temperature}

To keep the film from freezing, the radiative cooling must be compensated by
an influx of solar energy. The solar radiation flux at the earth orbit is
$J=1.37\times10^{3}%
%TCIMACRO{\unit{W}}%
%BeginExpansion
\operatorname{W}%
%EndExpansion
/%
%TCIMACRO{\unit{m}}%
%BeginExpansion
\operatorname{m}%
%EndExpansion
^{2}$ \cite{28}. Thus using the blackbody approximation and applying the
Stefan-Boltzmann law $J=\sigma_{0}T^{4}$, we find the equilibrium film
temperature $T=394%
%TCIMACRO{\unit{K}}%
%BeginExpansion
\operatorname{K}%
%EndExpansion
$.

The film must be transparent in order to be warmed uniformly. Perfect
absorption cannot be realized in transparent media, however. The equilibrium
temperature thus must be lower than $394%
%TCIMACRO{\unit{K}}%
%BeginExpansion
\operatorname{K}%
%EndExpansion
$. If this temperature is too low to maintain the fluidity, we can readily
raise it by dispersing absorbing particles such as carbon black in the
transparent liquid. By adjusting the volume percentage of these absorbing
particles, the film temperature can be kept in a range up to the blackbody
temperature $T$. Our assumed baseline temperature $T_{0}=298%
%TCIMACRO{\unit{K}}%
%BeginExpansion
\operatorname{K}%
%EndExpansion
$ is thus feasible.

Constancy of solar flux is important. If the sunlight is interrupted, the film
cools rapidly. There are two time scales associated with the cooling process.
One is the thermal equilibration time $t_{1}$. It is the time for the film to
reach thermal equilibrium via conduction. The other time scale $t_{2}$ is the
time to freeze via radiative cooling.

The time $t_{1}$ is approximately given by $h^{2}/\zeta$, where $h$ is the
film thickness and $\zeta$ is the thermal diffusivity. It is of the order of
$10^{-6}%
%TCIMACRO{\unit{m}}%
%BeginExpansion
\operatorname{m}%
%EndExpansion
^{2}/%
%TCIMACRO{\unit{s}}%
%BeginExpansion
\operatorname{s}%
%EndExpansion
$ for typical liquids \cite{0000}. We thus find%
\begin{equation}
t_{1}\sim10^{-6}%
%TCIMACRO{\unit{s}}%
%BeginExpansion
\operatorname{s}%
%EndExpansion
\text{.} \label{7}%
\end{equation}

To estimate $t_{2}$, we use the blackbody emission
approximation.\footnote{This is not a bad approximation. For instance, the
emissivity of water is about $1$ across thermal infrared region \cite{0001}.}
The balance of heat flux demands%
\begin{equation}
cAhdT=-2AJdt, \label{8}%
\end{equation}
where $J$ is the radiative flux $\sigma_{0}T^{4}$ as dictated by the
Stefan-Boltzmann law. $c$ is the specific heat and is of the order of $10^{7}%
%TCIMACRO{\unit{J}}%
%BeginExpansion
\operatorname{J}%
%EndExpansion
/%
%TCIMACRO{\unit{K}}%
%BeginExpansion
\operatorname{K}%
%EndExpansion
\cdot%
%TCIMACRO{\unit{m}}%
%BeginExpansion
\operatorname{m}%
%EndExpansion
^{3}$ for normal liquids \cite{0000}. $A$ is the surface area, and $h$ is the
film thickness. If the film temperature decreases from $T_{2}$ to $T_{1}$, we
find via integration
\begin{equation}
t_{2}=\frac{ch}{6\sigma_{0}}(\frac{1}{T_{1}^{3}}-\frac{1}{T_{2}^{3}}).
\label{9}%
\end{equation}
Our candidate liquid freezes at about $200%
%TCIMACRO{\unit{K}}%
%BeginExpansion
\operatorname{K}%
%EndExpansion
$ \cite{29}. We thus take $T_{1}=200%
%TCIMACRO{\unit{K}}%
%BeginExpansion
\operatorname{K}%
%EndExpansion
$ and $T_{2}=300%
%TCIMACRO{\unit{K}}%
%BeginExpansion
\operatorname{K}%
%EndExpansion
$, and it follows from Eq. (\ref{9}) that%
\begin{equation}
t_{2}\lesssim1%
%TCIMACRO{\unit{s}}%
%BeginExpansion
\operatorname{s}%
%EndExpansion
. \label{10}%
\end{equation}

The system thus rapidly reaches equilibrium via conduction [Eq. (\ref{7})] and
freezes within a few seconds [Eq. (\ref{10})] if the incoming solar radiation
flux is blocked. A\ frozen part could make the film shatter and should be
avoided. Thus the orbit of the system should be designed in such a way that
the film is kept constantly in the sunlight.

\subsection{Vapor pressure}

The lifetime $\tau$ of a liquid film against evaporation is determined by the
film thickness $h$, the evaporation mass flux $\phi$ (mass per unit area per
unit time), and the liquid mass density $\rho$ through the relation%
\begin{equation}
\tau=\frac{\rho h}{\phi}. \label{300}%
\end{equation}
For a micron-thick film with $\rho=10^{3}%
%TCIMACRO{\unit{kg}}%
%BeginExpansion
\operatorname{kg}%
%EndExpansion
/%
%TCIMACRO{\unit{m}}%
%BeginExpansion
\operatorname{m}%
%EndExpansion
^{3}$, a lifetime of at least $1$ year thus requires
\begin{equation}
\phi\leq10^{-10}%
%TCIMACRO{\unit{kg}}%
%BeginExpansion
\operatorname{kg}%
%EndExpansion
/%
%TCIMACRO{\unit{m}}%
%BeginExpansion
\operatorname{m}%
%EndExpansion
^{2}%
%TCIMACRO{\unit{s}}%
%BeginExpansion
\operatorname{s}%
%EndExpansion
. \label{304}%
\end{equation}
Thus a film of a conventional liquid with the designed geometry would
evaporate in seconds in the space environment. Proper liquids need to be
chosen to achieve such a small evaporation flux.

The evaporation flux $\phi$ of a liquid is related to its vapor pressure $p$
through the Langmuir formula \cite{16}:%
\begin{equation}
\phi=p\sqrt{\frac{m}{2\pi RT}}, \label{303}%
\end{equation}
where $m$ is the molar molecular weight and $R=8.31%
%TCIMACRO{\unit{J}}%
%BeginExpansion
\operatorname{J}%
%EndExpansion
/%
%TCIMACRO{\unit{mol}}%
%BeginExpansion
\operatorname{mol}%
%EndExpansion
\cdot%
%TCIMACRO{\unit{K}}%
%BeginExpansion
\operatorname{K}%
%EndExpansion
$.

\textit{Dow Corning }705 diffusion pump oil has typical product properties
$m=0.546%
%TCIMACRO{\unit{kg}}%
%BeginExpansion
\operatorname{kg}%
%EndExpansion
/%
%TCIMACRO{\unit{mol}}%
%BeginExpansion
\operatorname{mol}%
%EndExpansion
$ and $p=3\times10^{-10}%
%TCIMACRO{\unit{torr}}%
%BeginExpansion
\operatorname{torr}%
%EndExpansion
$ at $298%
%TCIMACRO{\unit{K}}%
%BeginExpansion
\operatorname{K}%
%EndExpansion
$ as introduced in Appendix A. Using these values, we infer via Eq.
(\ref{303}) that at the baseline temperature $298%
%TCIMACRO{\unit{K}}%
%BeginExpansion
\operatorname{K}%
%EndExpansion
$ the evaporation flux is $2\times10^{-10}%
%TCIMACRO{\unit{kg}}%
%BeginExpansion
\operatorname{kg}%
%EndExpansion
/%
%TCIMACRO{\unit{m}}%
%BeginExpansion
\operatorname{m}%
%EndExpansion
^{2}%
%TCIMACRO{\unit{s}}%
%BeginExpansion
\operatorname{s}%
%EndExpansion
$, which is larger than the condition Eq. (\ref{304}). Further manipulations
are therefore needed in order to reduce the vapor pressure and hence the
evaporation flux.

Vapor pressure can be reduced by lowering temperature. Using the vapor
pressure-temperature relation Eq. (\ref{B1}) for DC 705 fluid, we find that
the condition Eq. (\ref{304}) is satisfied at about $273%
%TCIMACRO{\unit{K}}%
%BeginExpansion
\operatorname{K}%
%EndExpansion
$. However, this reduced temperature increases film viscosity, which is an
unwelcome property. We discuss this issue in Appendix A.

Alternatively we can also achieve smaller evaporation flux with larger
molecular weight in the same class of molecules. The dependence of flux $\phi$
on molecular weight follows from the Eyring kinetics \cite{0002}:%
\begin{equation}
\phi\approx\frac{\rho_{s}}{t_{a}}\exp(-\frac{U}{k_{b}T}), \label{301}%
\end{equation}
where $\rho_{s}$ is the surface density of molecules, $1/t_{a}$ is the attempt
rate, and $U$ is the binding energy of a molecule to the liquid, which is
proportional to molecular weight. For polymers with $N$ repeat units, $U$ is
roughly proportional to $N$ and thus%
\begin{equation}
\phi\approx\frac{\rho_{s}}{t_{a}}\exp(-\alpha N), \label{302}%
\end{equation}
where $\alpha$ is a numerical factor and depends on temperature. Our candidate
liquid is such a polymer with $N=3$.

In the same class of molecules as our candidate liquid, we now estimate the
proper $N$ value that yields the requisite flux $\phi$ [Eq. (\ref{304})] at
$T_{0}=298%
%TCIMACRO{\unit{K}}%
%BeginExpansion
\operatorname{K}%
%EndExpansion
$, assuming $\rho_{s}/t_{a}$ weakly depends on molecular weight. Combining
Eqs. (\ref{303}) and (\ref{301}), we have%
\begin{equation}
p=\frac{\rho_{s}}{t_{a}}\sqrt{\frac{2\pi RT}{m}}\exp(-\frac{U}{k_{b}T}).
\label{305}%
\end{equation}
Using available data for DC 705 fluid \cite{29} and Eqs. (\ref{302}) and
(\ref{305}), we find at $T_{0}=298%
%TCIMACRO{\unit{K}}%
%BeginExpansion
\operatorname{K}%
%EndExpansion
$ the requisite evaporation flux Eq. (\ref{304}) can be satisfied for:
\begin{equation}
N\geq4. \label{306}%
\end{equation}
Eq. (\ref{306}), in connection with Eqs. (\ref{304}) and (\ref{303}), leads to%
\begin{equation}
p<10^{-11}%
%TCIMACRO{\unit{torr}}%
%BeginExpansion
\operatorname{torr}%
%EndExpansion
. \label{307}%
\end{equation}
\ \ \ \ \ \ \ \ \ \ 

With modest increment in molecular weight, one can thus achieve significantly
smaller evaporation flux and realize the designed lifetime of a year against
evaporation in the space environment.

\section{Stability of the thin liquid film against spontaneous rupture}

Thermal fluctuations in thin liquid films eventually cause them to collapse.
This important process has been studied in detail \cite{7,712,71,710,72}, and
the predicted lifetime against spontaneous rupture has been experimentally
confirmed \cite{70, 8}.

In the following we use the analysis of Sharma and Ruckenstein \cite{7,701}
developed for soap films to estimate the lifetime of the space film against
spontaneous rupture. This work treats both single component films and films
whose surfaces are saturated with insoluble surfactants. The stability of the
film depends on thickness $h$, dynamic viscosity $\eta$ of the film liquid,
surface tension $\sigma$, and surface concentration $\Gamma$ (number of
surfactant molecules per unit area) of surfactants. Although surface tension
helps to stabilize the film, the long range force, i.e., the van de Waals
disjoining pressure, tends to destabilize and eventually rupture it. When $h$
is small, typically a few micrometers, one can employ finite-amplitude
perturbative analysis to find the time constant $\tau$ of the fastest growing
wavevector $k_{m}$ of sinusoidal film surface modulations \cite{7}.

In the case of free thin films devoid of surfactants, one finds that the
fastest growing wavevector goes to zero, and%

\begin{equation}
\frac{1}{\tau}=\omega_{0}=\frac{A}{4\pi\eta h^{3}}, \label{01}%
\end{equation}
where $A$ is the Hamaker constant which characterizes the van der Waals
potential. It is of the order of $10^{-20}%
%TCIMACRO{\unit{J}}%
%BeginExpansion
\operatorname{J}%
%EndExpansion
$ for silicone oils like our candidate liquid.

In the case of free thin films with excess insoluble surfactants, one finds%
\begin{equation}
k_{m}=\frac{1}{h}\sqrt{\frac{A}{2\pi\sigma h^{2}}}, \label{02}%
\end{equation}
and%
\begin{equation}
\frac{1}{\tau}=\omega_{1}=\frac{A^{2}}{96\pi^{2}\eta\sigma h^{5}}%
=\frac{\left(  k_{m}h\right)  ^{2}}{12}\omega_{0}. \label{2}%
\end{equation}

Using the tabulated data for DC 705 oil, for the surfactant-free case we find
\begin{equation}
\tau\sim10^{2}%
%TCIMACRO{\unit{s}}%
%BeginExpansion
\operatorname{s}%
%EndExpansion
, \label{3}%
\end{equation}
and for the case with excess surfactants,%
\begin{equation}
\tau\sim10^{9}%
%TCIMACRO{\unit{s}}%
%BeginExpansion
\operatorname{s}%
%EndExpansion
\approx100%
%TCIMACRO{\unit{y}}%
%BeginExpansion
\operatorname{y}%
%EndExpansion
, \label{4}%
\end{equation}%
\begin{equation}
k_{m}\sim10^{2}\text{ }%
%TCIMACRO{\unit{m}}%
%BeginExpansion
\operatorname{m}%
%EndExpansion
^{-1}. \label{04}%
\end{equation}
The estimate Eq. (\ref{3}) makes the surfactant-free film infeasible whereas
the case with excess surfactants [Eq. (\ref{4})] yields more-than-adequate
lifetime. Compatible surfactants are therefore necessary for the space film to
achieve the designed lifetime against spontaneous rupture.

Experiments \cite{9} have confirmed that excess surfactant concentration
decreases the dominant perturbation wavelength and frequency, and thus
stabilizes thin liquid films. Rupture requires thinning of the initial film to
molecular dimensions. This thinning requires outward flow of the film away
from the thinnest region. Without surfactants, the fluid velocity is nearly
uniform throughout the thickness of the film. But an insoluble surfactant
layer inhibits flow at the surface, due to the viscoelastic effect of the
surfactant, so that the outward fluid velocity must nearly vanish at the
surface. Instead of being uniform throughout the thickness, the velocity
profile is therefore a parabolic Poiseuille flow. Thus a given flux of fluid
and a given thinning rate would require a much greater shear rate and much
greater dissipation when the surfactants are present. Since the rate of
dissipation is limited by the gain in van der Waals energy, the
surfactant-coated film thins more slowly than the uncoated film. Moreover
since aggregation of surface-active agents modifies the local surface tension,
surfactants can reduce the thinning flow by creating an opposing gradient in
surface tension as explained by the conventional Gibbs-Marangoni theory
\cite{901}.

The above analysis assumes that the surfactants are insoluble and surfactant
molecules are tightly packed to form a monolayer at each surface with the
liquid film in between. If surfactants are not completely insoluble, there
will be diffusion of surfactants in the bulk of the film and exchange of
surfactants between the bulk and the surfaces. These processes may reduce the
film stability. Theoretical and numerical studies \cite{1, 720} show a weak
dependence of the rupture time on solubility of surfactants but a strong
dependence on the surface viscosity $\eta_{s}$ that characterizes the
viscoelastic stress in surfactant monolayers. In addition, the rupture time
also depends strongly on the Marangoni parameter which characterizes how
effectively the surfactant concentration can modify surface tension.

The existence of appropriate surfactants for the unconventional liquid of our
film should not be taken for granted. The pump oils that provide needed
properties of liquidity and nonvolatility also have low surface tension. This
low initial surface tension limits the scope of compatible surfactants. In
order to be effective, surfactants must segregate to the surface, and this
must lower its surface tension significantly. However, it may not be easy to
find surfactants that can reduce the surface tension below that of the pump
oil we have chosen. In addition, the surfactant monolayers need to be strong
enough to yield large viscoelastic effect to ensure the film stability. Some
possible candidate surfactants for the pump oil are those used in oil foaming
\cite{721}, where surfactants are crucial to stabilize the microscopic plateau
films between adjacent oil bubbles.

The stability model we have referred to essentially assumes infinite lateral
scale, while in actual implementation the space film must be laterally finite
and supported by a frame. For a conventional wetting frame, the negative
Laplace pressure difference at the solid-liquid interface may drain the liquid
toward the frame. The resulting thinning region immediately near to the frame
may break and hence limit the film lifetime. Therefore the frame must be
engineered by choosing surface wetting properties and by adjusting local
interface geometry \cite{722} to counter the draining effect. For example, a
possible optimal frame can have a wedge-like boundary with an extending thin
edge towards the film. The sides of the edge can be rendered nonwetting while
the tip of the edge allows maximal wetting. A frame like this can achieve a
positive film surface curvature at the boundary and hence a positive Laplace
pressure difference. Thus the draining effect due to the frame can be overcome
though further experimental tests are needed. On the other hand, the possible
thinning near the frame boundary suggests the \textquotedblleft black
film\textquotedblright\ configuration, and we shall revisit it in the
Discussion section.

The case of excess insoluble surfactants yields more-than-adequate lifetime,
and this leaves room for us to further adjust film viscosity and thickness for
experimental purposes. For instance, by lowering viscosity we can improve
fluidity and thus enhance the dynamics in the film. Optimal viscosity and film
thickness should be decided in connection with other properties of the space
film that we shall consider later.

\section{External impacts in the space environment}

There are both potential chemical and physical impacts and damages to the film
in the space environment.

The potential chemical damage comes from radiation; this can degrade the film
in two different ways \cite{091}. Radiation can ionize oxygen and
double-bonded atoms in the environment. The resultant ions and free radicals
can interact with film molecules and damage the film. In space this damage can
be effectively reduced by avoiding low-earth orbit and therefore drastically
reducing these ions and free radicals. Radiation can also ionize film
molecules directly and degrade the film. Materials with sufficient resistance
to radiation are therefore required in order to reduce this aspect of damage.
Because of its high phenyl content, DC 705 pump oil has ideal resistance to
radiation among all organic compounds,\footnote{Energy-delocalizing aromatic
structural groups increase polymer stability by distributing energies of
excited states. Substituents on aromatic groups that extend the delocalized
bonding network are further stabilizers \cite{090}.} and it is stable in the
radiative space environment \cite{29}. Radiation damage thus poses no problem
for the space film based on such a liquid.

Micrometeoroids, however, may damage the film via physical impacts. Incoming
meteoroids affect the film both mechanically and thermally. These particles
may punch holes in the film directly by knocking out columns of film liquid.
They may also sail through or stay in the film, depositing in it most of their
energy and exciting elastic waves. At the same time, part of the kinetic
energy of these particles will be transferred to the film in the form of
thermal energy. This process will increase local temperature and consequently
change the local properties such as surface tension and viscosity. It can even
evaporate film liquid in the impact regions and produce high pressure gas in
the holes. We must assess these potential physical damages and propose
possible ways to reduce them.

\subsection{meteoroid-induced hole nucleation in the film}

Since both mechanical and thermal effects of meteoroid impacts generally
produce holes in the film, we now first focus on a single hole and analyze how
it forms, evolves, and affects the film.

If a hole in the film is too large, it will be unstable and will continue to
expand \cite{092}. We study the stability of a dewetting hole of radius $r$ by
considering the change of free energy $\Delta F\left(  r\right)  $ associated
with the formation of the hole. We ignore long range forces and consider only
surface energy. We assume the shape of the hole to be semicircular in cross
section, and $r\ll L$. The change of free energy is
\begin{equation}
\Delta F\left(  r\right)  =\sigma(2\pi rh-2\pi r^{2}), \label{51}%
\end{equation}
and
\begin{equation}
\frac{d\Delta F}{dr}=2\pi\sigma(h-2r). \label{52}%
\end{equation}
The condition $d\Delta F/dr=0$ demands $r=h/2$ which is the critical size of a
dewetting hole. If the radius of the hole is smaller than $h/2$, $d\Delta
F/dr>0$, and the hole will shrink and eventually disappear. Otherwise $d\Delta
F/dr<0$, and the hole will expand and potentially damage the film.

It was found experimentally \cite{010} that dewetting holes expand at a
constant velocity.\footnote{For very viscous liquid films, with viscosity up
to $10^{6}$cSt, the hole radius grows exponentially with time \cite{010}.} The
classical Dupre-Culick law \cite{09} provides simple estimate of the expansion
velocity $V$. The dewetting hole is surrounded by a rim collecting the liquid,
where the dynamics localizes. The change of momentum per unit length of the
rim satisfies%
\begin{equation}
\frac{dP}{dt}=V\frac{dm}{dt}=2\sigma, \label{5201}%
\end{equation}
where $m$ is the mass of the rim per unit length. Since $dm/dt=\rho hV$ where
$\rho$ is liquid density, we can solve for $V$:%
\begin{equation}
V=\sqrt{\frac{2\sigma}{\rho h}}\approx1%
%TCIMACRO{\unit{m}}%
%BeginExpansion
\operatorname{m}%
%EndExpansion
/%
%TCIMACRO{\unit{s}}%
%BeginExpansion
\operatorname{s}%
%EndExpansion
. \label{5202}%
\end{equation}
The expanding hole can reach the boundary of the film within an hour.

Thus a hole with size of order $h$ or greater will expand and eventually
destroy the film and must be avoided. If an incoming meteoroid of size $r$ can
produce a hole of size $h$ in the film, a necessary condition is that its
kinetic energy $K=2/3\pi r^{3}\rho_{m}v^{2}$ must overcome the surface energy
$2\pi\sigma h^{2}$, \textit{i.e.},%
\begin{equation}
r>r_{c}=\left(  \frac{3\sigma h^{2}}{\rho_{m}v^{2}}\right)  ^{1/3}%
\approx10^{-9}%
%TCIMACRO{\unit{m}}%
%BeginExpansion
\operatorname{m}%
%EndExpansion
, \label{549}%
\end{equation}
where the meteoroid density $\rho_{m}$ is about $5\times10^{3}%
%TCIMACRO{\unit{kg}}%
%BeginExpansion
\operatorname{kg}%
%EndExpansion
/%
%TCIMACRO{\unit{m}}%
%BeginExpansion
\operatorname{m}%
%EndExpansion
^{3}$, and the incoming speed $v$ is as high as $30%
%TCIMACRO{\unit{km}}%
%BeginExpansion
\operatorname{km}%
%EndExpansion
/%
%TCIMACRO{\unit{s} }%
%BeginExpansion
\operatorname{s}
%EndExpansion
$.\footnote{We use the mean orbital velocity of the earth $29.78%
%TCIMACRO{\unit{km}}%
%BeginExpansion
\operatorname{km}%
%EndExpansion
/%
%TCIMACRO{\unit{s}}%
%BeginExpansion
\operatorname{s}%
%EndExpansion
$ for the estimate.} $r_{c}$ thus sets a lower bound for the size of incoming
meteoroids that are dangerous to the film.

Eq. (\ref{549}) also shows that for an incoming meteoroid with size comparable
to $h$ or greater, its kinetic energy is much larger than the corresponding
surface energy of the impact area, and the surface tension can be ignored
during the impact. Consequently a meteoroid of such a size will simply knock
out the corresponding column of the film liquid and produce an expanding hole.

Thus incoming meteoroids with sizes larger than $h=10^{-6}%
%TCIMACRO{\unit{m}}%
%BeginExpansion
\operatorname{m}%
%EndExpansion
$ are dangerous to the film while meteoroids with size smaller than
$r_{c}=10^{-9}%
%TCIMACRO{\unit{m}}%
%BeginExpansion
\operatorname{m}%
%EndExpansion
$ are harmless. More detailed studies of the interaction process are needed
with respect to sizes $r_{c}<r<h$.

\subsection{interactions between meteoroids and the film}

Meteoroids interact with the film both mechanically and thermally. They can
directly impart momentum to the film during collisions, and the momentum
propagates in the film via elastic waves or shock fronts \cite{009}. They can
also boil the local film liquid, and the resultant high-pressure gas transfers
its energy to the film via rapid expansion. At the same time viscous
dissipation reduces the kinetic energy to thermal energy. The total effect of
the impact is therefore the result of the competing propagation process and
dissipation process.

The detailed interaction thus depends on many physical properties and is a
very complicated problem in engineering. First-principles estimation of this
process goes beyond the scope of the paper. To estimate the impact effect, we
instead consider available empirical information.

In the formation of craters due to high-speed meteorite impact on an
astronomical body, an empirical relation between the diameter $d$ of a crater
and the total kinetic energy $K$ of the incoming meteorite is given by
\cite{40}:%
\begin{equation}
d\propto K^{1/3}. \label{570}%
\end{equation}
The proportionality coefficient is of the order of unity in SI units.

If we apply the relation Eq. (\ref{570}) to the formation of holes in the
space film and insert the expression for the kinetic energy $K=2/3\pi
r^{3}\rho_{m}v^{2}$, we find%
\begin{equation}
D\approx10r, \label{571}%
\end{equation}
where $D$ is the diameter of the hole and $r$ is the size of the meteoroid.
Thus a meteoroid with size $0.1h=10^{-7}%
%TCIMACRO{\unit{m}}%
%BeginExpansion
\operatorname{m}%
%EndExpansion
$ can produce a hole of size $h$ and should be avoided, while meteoroids with
smaller sizes would not destroy it according to this criterion.

The hole size based on relation Eq. (\ref{570}) could be overestimated. The
cratering process gives only a rough guide to the expected hole size $D$.
Craters form at the surface of a bulk material rather than a thin layer. Thus
the confinement of energy and stress is likely greater in the cratering
geometry. Moreover the solid material forming a crater is stronger than the
liquid material of our film. Both these effects would be expected to increase
the retarding forces on the projectile. Furthermore, in forming the crater,
the incoming meteorite fragments and cannot pass deeply into the solid medium.
Most of its initial energy is thus released near to the impact surface, and
there is severe jetting. All these effects lead to greater disturbance in a
smaller volume. It is thus natural that greater damage is done in the catering
geometry so that the hole diameter would be expected to be larger there.

High-speed impact has also been studied using silica aerogel as the target.
Aerogel is a highly porous solid material with a density of $50%
%TCIMACRO{\unit{kg}}%
%BeginExpansion
\operatorname{kg}%
%EndExpansion
/%
%TCIMACRO{\unit{m}}%
%BeginExpansion
\operatorname{m}%
%EndExpansion
^{3}$ \cite{53}, and it has many exceptional physical and chemical properties.
Both simulations based on fired projectiles and actual results from meteoroids
trapped in aerogel suggest that high-speed meteoroids can penetrate the
aerogel to a depth of an order of $10$ to $100$ times the size of the
meteoroids, without severe heating or fragmentation \cite{50}. The typical
impacts produce carrot-shaped tracks that begin with entry holes one order of
magnitude wider than the diameters of incoming meteoroids. There are no
obvious fracturing near the tracks.

Different and independent experimental evidences thus lead to the same
empirical relation Eq. (\ref{571}). The solid medium where craters form is
stronger than the liquid film while the aerogel, with density of $1/20$ of
that of DC\ 705 liquid, is much weaker. Interactions with high-speed
projectiles can be qualitatively different in different media with different
geometries. Our findings, however, suggest that the empirical relation between
the hole size and the diameter of the incoming projectile is robust in a broad
spectrum of impact medium properties. It is thus feasible to assume the same
relation in the case of space film, \textit{i.e.}, the incoming meteoroid with
size $0.1h$ will produce a hole of size $h$ and should be avoided. Exact
results with respect to meteoroid size should be determined by further
experimental tests in the actual liquid environment.

\subsection{meteoroid flux and mass concentration}

The flux of meteoroids (number of impacts per unit area per unit time) of mass
$m$ and greater satisfies the empirical power law \cite{12}:%
\begin{equation}
n(m)=n_{0}\left(  \frac{m}{m_{0}}\right)  ^{\alpha}, \label{055}%
\end{equation}
where $\alpha$ is the mass distribution index, $m_{0}$ is a constant with
dimension of mass, and $n_{0}$ is the flux of mass $m_{0}$ and greater.

The meteoroid flux is largest near the earth orbit, and further away from the
earth the interplanetary particle flux is reduced \cite{10}. Earlier
measurements \cite{13} showed that $\alpha$ ranges between $-0.27$ and $-0.34$
at mass threshold of $10^{-9}$ $%
%TCIMACRO{\unit{kg}}%
%BeginExpansion
\operatorname{kg}%
%EndExpansion
$ in the space region between $1$ and $1.6$ AU. At mass threshold of
$10^{-12}$ $%
%TCIMACRO{\unit{kg}}%
%BeginExpansion
\operatorname{kg}%
%EndExpansion
$, measurements by Pioneer 11 showed that $\alpha$ is $-0.5$ between $2$ and
$5 $ AU \cite{14}.

Moreover at mass threshold of $10^{-12}$ $%
%TCIMACRO{\unit{kg}}%
%BeginExpansion
\operatorname{kg}%
%EndExpansion
$, the detectors on Pioneer 10 and 11 recorded an almost constant penetration
flux of $10^{-6}$ impacts$/%
%TCIMACRO{\unit{m}}%
%BeginExpansion
\operatorname{m}%
%EndExpansion
^{2}%
%TCIMACRO{\unit{s}}%
%BeginExpansion
\operatorname{s}%
%EndExpansion
$ in the space region between $1$ and $18$ AU \cite{11}. At mass threshold of
$10^{-16}%
%TCIMACRO{\unit{kg}}%
%BeginExpansion
\operatorname{kg}%
%EndExpansion
$, detectors on Galileo and Ulysses recorded a penetration flux of $10^{-4}$
impacts$/%
%TCIMACRO{\unit{m}}%
%BeginExpansion
\operatorname{m}%
%EndExpansion
^{2}%
%TCIMACRO{\unit{s}}%
%BeginExpansion
\operatorname{s}%
%EndExpansion
$ in the same region \cite{12}. We infer from these data that $\alpha=-0.5$,
which is consistent with direct measurements.

We have shown that high-speed meteoroids with sizes of $0.1h$ or greater can
produce expanding holes directly and must be avoided. A meteoroid with the
critical size $0.1h$ has a mass of $10^{-17}%
%TCIMACRO{\unit{kg}}%
%BeginExpansion
\operatorname{kg}%
%EndExpansion
$. With such a mass threshold, previous data and Eq. (\ref{055}) show a
penetration flux of $10^{-4}$ impacts$/%
%TCIMACRO{\unit{m}}%
%BeginExpansion
\operatorname{m}%
%EndExpansion
^{2}%
%TCIMACRO{\unit{s}}%
%BeginExpansion
\operatorname{s}%
%EndExpansion
$. Thus there will be $100$ impacts per second on the space film with surface
area of $1%
%TCIMACRO{\unit{km}}%
%BeginExpansion
\operatorname{km}%
%EndExpansion
^{2}$, implying a very short film lifetime. On the other hand, to achieve the
desired lifetime of a few years, i.e., $10^{7}%
%TCIMACRO{\unit{s}}%
%BeginExpansion
\operatorname{s}%
%EndExpansion
$, requires a flux of $10^{-13} $ impacts/$%
%TCIMACRO{\unit{m}}%
%BeginExpansion
\operatorname{m}%
%EndExpansion
^{2}%
%TCIMACRO{\unit{s}}%
%BeginExpansion
\operatorname{s}%
%EndExpansion
$ or less that corresponds to a mass threshold $m\gtrsim100%
%TCIMACRO{\unit{kg}}%
%BeginExpansion
\operatorname{kg}%
%EndExpansion
$. To survive impacts up to this mass threshold would require $h\approx1%
%TCIMACRO{\unit{m}}%
%BeginExpansion
\operatorname{m}%
%EndExpansion
$, which is clearly not feasible.

Thus additional manipulations must be employed to protect the film against
impacts from meteoroids with size of $0.1h$ or greater. Since interplanetary
dust velocities are assumed to be mainly concentrated in certain directions
within the ecliptic plane of the solar system \cite{15}, one could possibly
avoid major impacts by tilting the film to be tangent to the predominant
meteoroid orbits and using a shield. One can also imagine modifying the film
liquid so that a hole does not lead to catastrophic failure. We sketch some
ideas in the Discussion section below. In what follows, we shall suppose that
some means of protecting the film against rupture by micrometeoroids has been employed.

Next we assess the film mass loss due to bombardments from incoming
meteoroids. We rewrite the mass flux Eq. (\ref{055}) as the flux with respect
to size $r$ and greater with $\alpha=-0.5$:
\begin{equation}
n(r)=n_{0}\left(  \frac{r}{r_{0}}\right)  ^{-3/2}, \label{547}%
\end{equation}
where $r_{0}$ is the size of the meteoroid with mass $m_{0}$. The flux of
meteoroids between size $r$ and $r+dr$ is then $-\left(  dn/dr\right)
dr=n_{0}(r/r_{0})^{-5/2}dr/r_{0}$. If each incoming meteoroid of size $r$
knocks out a column of film liquid with area $\pi(10r)^{2}$, then the total
mass loss flux due to bombardments from meteoroids up to size $r_{M}$ is
\begin{align}
\phi_{M}  &  \approx\rho h\int_{0}^{r_{M}}100\pi r^{2}\left(  -\left(
dn/dr\right)  \right)  dr\nonumber\\
&  =200\pi\rho hr_{M}^{2}n\left(  r_{M}\right)  . \label{600}%
\end{align}
If we set the cut-off size $r_{M}$ to be $1%
%TCIMACRO{\unit{m}}%
%BeginExpansion
\operatorname{m}%
%EndExpansion
$ which, as shown above, corresponds to $1$ impact per year on the entire
film, we find $\phi_{M}$ $\approx10^{-13}%
%TCIMACRO{\unit{kg}}%
%BeginExpansion
\operatorname{kg}%
%EndExpansion
/%
%TCIMACRO{\unit{m}}%
%BeginExpansion
\operatorname{m}%
%EndExpansion
^{2}%
%TCIMACRO{\unit{s}}%
%BeginExpansion
\operatorname{s}%
%EndExpansion
$. $\phi_{M}$ is thus much smaller than the requisite evaporation flux $\phi$
[Eq. (\ref{304})], and the mass loss due to meteoroid bombardments is negligible.

\section{Two-dimensional hydrodynamics and turbulence}

Two-dimensional hydrodynamics and turbulence have attracted sustained
scientific attention \cite{19}: they are the basic model for geophysical and
planetary flow, and are relevant to the large-scale dynamics of ocean and
atmosphere, and are also applied to strongly magnetized plasmas.

Mathematically two-dimensional hydrodynamics is governed by the incompressible
Navier-Stokes equation%
\begin{equation}
\partial_{t}\mathbf{v}+\mathbf{v}\cdot\nabla\mathbf{v}=-\nabla p+\frac{\eta
}{\rho}\nabla^{2}\mathbf{v} \label{11}%
\end{equation}
for the velocity field $\mathbf{v}$, where $p$ is a pressure field chosen so
that $\nabla\cdot\mathbf{v=}0$, and $\rho$ is the density, and $\eta$ is the
dynamic viscosity. \textquotedblleft Two-dimensional
turbulence\textquotedblright\ corresponds to high Reynolds number solutions of
the Navier-Stokes equation that depend only on two Cartesian coordinates
\cite{20}. In this case, it is straightforward to verify that the component of
the velocity along the third coordinate axis satisfies an advection-diffusion
equation and decouples from the horizontal flow.

Both theoretical models \cite{224} and experimental measurements \cite{225}
show that two-dimensional turbulence has a unique \textquotedblleft double
cascades\textquotedblright\ structure: enstrophy (mean-square vorticity) is
transported \textit{downward} from the injection scale to the viscous scale,
while energy is transported \textit{upward} from the injection scale to a
larger scale imposed by the boundary of the system. The transfer of enstrophy
or energy is \textit{inertial }in these ranges without loss. The upward
cascade of energy consequently gives rise to mergers of vortices and hence the
emergence of large-scale coherent patterns compatible with the boundary where
the energy eventually dissipates. Thus two-dimensionality imposes additional
boundary conditions on the flow. This could be relevant to the geometry of our
proposed space film experiment.

For a film configured as a thin layer of base liquid between two layers of
surfactants, some modifications to the theory need to be introduced
\cite{221}. First, the pressure term in Eq. (\ref{11}) is incomplete. The
surface tension becomes more important as the film gets thinner. For a thin
film with thickness $h$, the pressure $p$ is replaced by the surface tension
term $\sigma/h$. Second, the surfactant layers modify the film viscosity in
such a way that they bring about additional dissipation that characterizes the
viscous friction in the film plane. This additional dissipative force can be
formulated as%
\begin{equation}
F_{s}=2\frac{\eta_{s}}{\rho h}\nabla^{2}\mathbf{v,} \label{12}%
\end{equation}
where $\rho$ is the liquid density, and $\eta_{s}$ is the surface viscosity of
the two surfactant layers. $\eta_{s}$ is determined by the nature of the
surfactant and film liquid, and it also depends on surfactant concentration.
In the case of soap films, $\eta_{s}/h$ is comparable to the dynamic viscosity
of water \cite{221}, but more general data for $\eta_{s}$ are not available.

The effective dynamic viscosity of a thin liquid film is therefore%

\begin{equation}
\eta_{f}=\eta+2\frac{\eta_{s}}{h}, \label{13}%
\end{equation}
where $\eta$ is the dynamic viscosity of the base liquid. The Reynolds number
for the two-dimensional flow in such a film is%
\begin{equation}
\operatorname{Re}=\frac{\rho Ul_{inj}}{\eta+2\eta_{s}/h}. \label{14}%
\end{equation}
where $l_{inj}$ is the energy injection length scale at which the system is driven.

We now assess the highest Reynolds number attainable for the space film by
estimating the nominal flow velocity $U$. Since the film is thin, the
stability of the film is dependent on surface forces. Pieces of the film have
a typical lateral velocity that must go to zero as the pieces approach the
boundary. This process could cause the accumulation of liquid in the local
region and variance in film thickness, and it could damage the film. Any
thickening of the film requires a potential flow such that $\nabla
\cdot\mathbf{v\neq}0$ as discussed after Eq. (\ref{04}). Such a flow causes
compression of the surfactant layer and a compressive stress $\Pi$ due to the
resulting gradient in surface concentration of surfactants. It is the
compressive stress that stops the flow towards the boundary. Since the stress
is determined by the nature of the surfactants and the film liquid, this
process limits the velocity in a range that is compatible with the film
stability. In addition, a restoring force is needed in order to prevent the
film from bulging significantly into the third dimension, and this force is
provided by surface tension.

For a film patch of unit area moving with the nominal velocity $U$, the
required decelerating force is
\begin{equation}
h\rho\frac{dU}{dt}=h\rho\frac{dU}{dx}\frac{dx}{dt}=h\rho\frac{d(U^{2}/2)}{dx}.
\label{15}%
\end{equation}
This force is provided by the gradient of the compressive stress $d\Pi/dx$,
and therefore%
\begin{equation}
\Delta\Pi\approx\frac{1}{2}h\rho U^{2}. \label{150}%
\end{equation}
We can estimate $\Delta\Pi$ by considering the difference in surface tension
with and without surfactants, i.e.,%
\begin{equation}
\Pi\sim\sigma-\sigma_{\Gamma}. \label{151}%
\end{equation}
Since $\Delta\Pi\lesssim\Pi$, $\Delta\Pi$ is thus smaller than $\sigma$. And
since $\sigma_{\Gamma}$ provides the restoring force preventing the film from
bulging, $\sigma_{\Gamma}$ cannot be too small. We can therefore assign
$\Delta\Pi$ as a finite fraction of $\sigma$ such as $1/2$, and thus
\begin{equation}
U\approx\sqrt{\frac{\sigma}{\rho h}}\sim1%
%TCIMACRO{\unit{m}}%
%BeginExpansion
\operatorname{m}%
%EndExpansion
/%
%TCIMACRO{\unit{s}}%
%BeginExpansion
\operatorname{s}%
%EndExpansion
. \label{16}%
\end{equation}
The highest Reynolds number for the film based on DC 705 liquid is%

\begin{equation}
\operatorname{Re}\approx\frac{\rho UL}{\eta}\sim10^{7}, \label{17}%
\end{equation}
where we have ignored the surface viscosity [Eq. (\ref{14})] and have chosen
$l_{inj}$ to be its maximum value, the system size $L$.\footnote{In many
experiments \cite{225} one chooses $l_{inj}\ll L$ so as to study the inverse
cascade of energy mentioned earlier in this section. Nevertheless the nominal
Reynolds number given in Eq. (\ref{17}) shows the large range of length scales
accessible to the space film experiment.}\ \ 

We have seen that the surfactants retard but do not eliminate divergence in
the two-dimensional flow, creating a variable thickness. These flows create
additional dissipation not accounted for by bulk or surface viscosity. This
dissipation could significantly reduce the Reynolds number though we are
unable to estimate it here. Moreover, the above analysis is based on
conventional thin-liquid-film model, and it may not be applied to other stable
film configurations such as the \textquotedblleft black film\textquotedblright%
\ that we shall discuss in the Discussion section.

In a normal laboratory environment, high pressure air in a wind
tunnel,\footnote{Princeton Gas Dynamics Lab SuperPipe Facility. The range of
Reynolds number achieved there is from $5\times10^{3}$ to $3.8\times10^{7}$.}
or fluids with extremely low viscosity, such as liquid helium \cite{222, 223},
are used to achieve high Reynolds numbers up to $10^{7}$. These methods only
apply to the three-dimensional turbulence, however. Our result Eq. (\ref{17})
is thus comparable to the highest Reynolds number achievable in normal laboratories.

Besides potentially achieving very high Reynolds number, the space film also
lasts for a comparatively long time. This feature can be very important for
the experimental study of turbulence. Turbulence can be divided into two types
\cite{200}: forced steady-state turbulence and decaying turbulence. For forced
turbulence the two-dimensional flow is constantly stirred by a stirring
device, and for decaying turbulence the system is stirred at the beginning,
and then the stirring device is removed. High Reynolds number is more
desirable in the first case since we are interested in properties of the
steady state. In the decaying turbulence case, the whole dynamics is
important, and the time scale needs to be long enough to allow the dynamics to
fully develop. In principle, both types of turbulence can be realized in the
space film.

\section{Comparison with other recent experiments on two-dimensional
hydrodynamics}

Current experimental studies of two-dimensional hydrodynamics and turbulence
are based on two different realizations \cite{200}. The first approach is to
use soap films to model two-dimensional flows. Turbulent flows are generated
by dragging thin objects along film surfaces \cite{221, 22}. The problem with
this method is that the air near the film surfaces affects the flow, and
additional viscous friction needs to be included in order to understand the
dynamics \cite{220}. The second approach is to use shallow layer of solution
to approximate the quasi two-dimensional flow \cite{230, 23}. Electromagnetic
forces are used to make the solution flow. The difficulty with this method is
that the depth of the fluid layer and the friction from the bottom of the cell
affect the dynamics \cite{24}. Therefore in the normal laboratory environment,
it is very difficult to completely decouple the two-dimensional flow with the
solid surfaces or the gaseous phase that surrounds it \cite{220}. The
attainable Reynolds number in current experimental studies of two-dimensional
turbulence is of the order of $10^{4} $ due to technical
restrictions.\footnote{In numercal studies of two-dimensional turbulence, the
highest attained Reynolds number is $10^{4}$ to $10^{5}$ \cite{60}.}

To compare with the space film, we focus on current experimental studies based
on soap films. Some important developments have been reported recently. In the
soap tunnel device \cite{25}, the maximum film speed is $30%
%TCIMACRO{\unit{cm}}%
%BeginExpansion
\operatorname{cm}%
%EndExpansion
/%
%TCIMACRO{\unit{s}}%
%BeginExpansion
\operatorname{s}%
%EndExpansion
$ for a film with thickness of $6%
%TCIMACRO{\unit{\U{3bc}m}}%
%BeginExpansion
\operatorname{\mu m}%
%EndExpansion
$. In the flowing film method \cite{26}, the whole film is falling, and new
fluid is supplied at the top. A maximum velocity of a few meters per second is
achieved. However since the film is falling fast, its width and thickness are
difficult to control. Most recently some improvements of the falling film
experiment have been made \cite{27}, and two-dimensional nonlinear dynamics
has been studied with improved experimental techniques. The flow speed in such
falling films ranges between $0.5$ and $4%
%TCIMACRO{\unit{m}}%
%BeginExpansion
\operatorname{m}%
%EndExpansion
/%
%TCIMACRO{\unit{s}}%
%BeginExpansion
\operatorname{s}%
%EndExpansion
$, with film thickness between $1$ and $10%
%TCIMACRO{\unit{\U{3bc}m}}%
%BeginExpansion
\operatorname{\mu m}%
%EndExpansion
$, and typical film size of $3%
%TCIMACRO{\unit{m}}%
%BeginExpansion
\operatorname{m}%
%EndExpansion
$ tall and $10%
%TCIMACRO{\unit{cm}}%
%BeginExpansion
\operatorname{cm}%
%EndExpansion
$ wide. Giant films, with size as large as $10%
%TCIMACRO{\unit{m}}%
%BeginExpansion
\operatorname{m}%
%EndExpansion
$, are also achievable with similar techniques. The Reynolds number for giant
films is about $10^{6}$. No experimental results on such giant films have been
reported, however.

The soap-film-based experiments have some common restrictions. First, as
mentioned before, the dragging force due to the air near the film can never be
totally eliminated, and this additional viscous friction affects the
two-dimensional nonlinear dynamics. Second, since static large films cannot
last long under normal laboratory conditions due to the drainage of gravity,
these films are necessarily falling vertically with base liquid perpetually
injected into the system. Two-dimensional nonlinear dynamic processes with
time scale longer than the characteristic time of the falling system cannot
fully develop in such films.

Compared to these realizations, the space film can potentially achieve
considerably higher Reynolds number, with much larger lateral dimension, and
last for a much longer time scale. External forces are needed in order to
accelerate the base liquid to its nominal velocity of $1%
%TCIMACRO{\unit{m}}%
%BeginExpansion
\operatorname{m}%
%EndExpansion
/%
%TCIMACRO{\unit{s}}%
%BeginExpansion
\operatorname{s}%
%EndExpansion
$, which is comparable to those typical velocities achieved in conventional
soap films.

\section{Forces on the space film}

\subsection{Support force}

The space film needs to be supported by a frame. The frame should support the
film tension and the force due to the base liquid flow: the frame must provide
a force to decelerate the liquid near the boundary. According to our previous
analysis, when the system achieves its nominal velocity, i.e., $U=1%
%TCIMACRO{\unit{m}}%
%BeginExpansion
\operatorname{m}%
%EndExpansion
/%
%TCIMACRO{\unit{s}}%
%BeginExpansion
\operatorname{s}%
%EndExpansion
$, these two forces are comparable to each other and can be estimated as:%
\begin{equation}
F=\sigma L=10%
%TCIMACRO{\unit{N}}%
%BeginExpansion
\operatorname{N}%
%EndExpansion
. \label{35}%
\end{equation}
A light frame can easily support such a small force. In addition, the frame
must also be strong enough against buckling.

\subsection{External forces}

The films encounters perturbing forces from the space environment. These
forces should not perturb the film too much. On the other hand, deliberate
forcing is needed in order to create the desired turbulent flow. We consider
each type of forces in turn.

\subsubsection{Environmental forces}

\textbf{Radiation pressure}.\textbf{\ }The solar radiation flux at the earth
orbit is $1.37\times10^{3}%
%TCIMACRO{\unit{W}}%
%BeginExpansion
\operatorname{W}%
%EndExpansion
/%
%TCIMACRO{\unit{m}}%
%BeginExpansion
\operatorname{m}%
%EndExpansion
^{2}$ \cite{28}. The corresponding momentum flux is $10^{-5}%
%TCIMACRO{\unit{Pa}}%
%BeginExpansion
\operatorname{Pa}%
%EndExpansion
$, and the total force on the film is no more than $10%
%TCIMACRO{\unit{N}}%
%BeginExpansion
\operatorname{N}%
%EndExpansion
$. This force is comparable to the characteristic dynamic force [Eq.
(\ref{35})]. \ \ \ \ \ \ 

\textbf{Solar wind pressure}.\textbf{\ }The solar wind pressure due to
incoming charged particles is of the order of $10^{-9}%
%TCIMACRO{\unit{Pa}}%
%BeginExpansion
\operatorname{Pa}%
%EndExpansion
$ \cite{28} and is negligible compared to the radiation pressure.

\textbf{Electrostatic force}.\textbf{\ }In the space environment the film is
charged, and it can interact with the ambient electric field. The incoming
charged particles can charge the film to as high as $10%
%TCIMACRO{\unit{V}}%
%BeginExpansion
\operatorname{V}%
%EndExpansion
$ \cite{150}. This finding is comparable to the approximation by using image
charge attraction (we assume the attraction potential to be that of two
elementary charges at an atomic separation). We can estimate the charge of the
film via the formula $V_{e}=\left(  1/4\pi\epsilon_{0}\right)  \left(
Q/L\right)  $. We set $V_{e}$ to be $10%
%TCIMACRO{\unit{V}}%
%BeginExpansion
\operatorname{V}%
%EndExpansion
$ and find $Q=10^{-6}%
%TCIMACRO{\unit{C}}%
%BeginExpansion
\operatorname{C}%
%EndExpansion
$. If we assume the moving electric field to be a few millivolt per meter that
is the characteristic field of the ionosphere, it gives a total force $10^{-9}%
%TCIMACRO{\unit{N}}%
%BeginExpansion
\operatorname{N}%
%EndExpansion
$. Again this force is very small and is negligible compared to other forces.

\textbf{Marangoni forces}.\textbf{\ }One can achieve a desired tangential
gradient of surface tension along the film surface via a nonuniform
temperature. The resultant Marangoni forces could be the driving force.
Nonuniformity in surfactant concentration can also yield variations in surface
tension. Moreover, external perturbations and hydrodynamic flow can induce
variations in thickness \cite{151}, which will introduce additional
Marangoni-type forces. Further study of Marangoni forces on the space film is
needed for the film design.

\subsubsection{Direct mechanical forces}

The forces needed to generate the turbulent flow can be supplied mechanically
or via electromagnetic forces. These forces should not be larger than the
characteristic force, i.e., $10%
%TCIMACRO{\unit{N}}%
%BeginExpansion
\operatorname{N}%
%EndExpansion
$. We may use existing soap-film experiments as a guide in implementing these
forces. We may, for example, insert a mechanical stirrer into the film
\cite{27}. The stirrer could be supported by wires suspended above the film.
It could be driven by conventional electric motors. The size and motion of the
stirrer determine the energy injection length scale $l_{inj}$. In the same
manner, the flow could be observed by conventional cameras also suspended
above the film. Detailed analysis and design of these forces involve technical
and engineering issues that go beyond the scope of this paper.

\section{Characteristic times}

Besides the lifetime and the previously discussed time to freeze, there are
other time scales that are relevant to the two-dimensional dynamics in the
film and are important for film design.

\textbf{Acceleration time}.\textbf{\ }The acceleration force should not be
larger than the characteristic force, i.e., $10%
%TCIMACRO{\unit{N}}%
%BeginExpansion
\operatorname{N}%
%EndExpansion
$, otherwise the surface tension will not be able to sustain the system. If
the acceleration force is $1%
%TCIMACRO{\unit{N}}%
%BeginExpansion
\operatorname{N}%
%EndExpansion
$ and the total mass of the film is $10^{3}%
%TCIMACRO{\unit{kg}}%
%BeginExpansion
\operatorname{kg}%
%EndExpansion
$, the acceleration will be $10^{-3}%
%TCIMACRO{\unit{m}}%
%BeginExpansion
\operatorname{m}%
%EndExpansion
/%
%TCIMACRO{\unit{s}}%
%BeginExpansion
\operatorname{s}%
%EndExpansion
^{2}$. It will take $10^{3}%
%TCIMACRO{\unit{s}}%
%BeginExpansion
\operatorname{s}%
%EndExpansion
$ to accelerate the system to its nominal velocity $U=1%
%TCIMACRO{\unit{m}}%
%BeginExpansion
\operatorname{m}%
%EndExpansion
/%
%TCIMACRO{\unit{s}}%
%BeginExpansion
\operatorname{s}%
%EndExpansion
$.

\textbf{Transit time}.\textbf{\ }The transit time is $L/U\approx10^{3}%
%TCIMACRO{\unit{s}}%
%BeginExpansion
\operatorname{s}%
%EndExpansion
$ for the liquid to flow from one side to the other side of the film. The
experiment needs to last many transit times for the film to reach the steady
state. The transit time provides the lower bound for the film lifetime. Our
original designed lifetime of $1$ year readily satisfies this condition.

\textbf{Relaxation time}. In connection to the double-cascade structure,
two-dimensional turbulence first relaxes to a metaequilibrium state with the
emergence of large-scale coherent patterns and little energy loss, and the
metaequilibrium state then decays through the energy dissipation at the
boundary \cite{153, 152}. In the decay of the metaequilibrium state, we assume
the shear rate to be of order $U/L$. The total power dissipated by the viscous
force is thus $P_{\eta}\approx\eta U\left(  U/L\right)  Lh=\eta U^{2}%
h\approx10^{-5}%
%TCIMACRO{\unit{W}}%
%BeginExpansion
\operatorname{W}%
%EndExpansion
$. The viscous relaxation time is defined to be the total kinetic energy of
the film divided by the total power due to the viscous force:
\begin{equation}
\tau_{\eta}=\frac{K}{P_{\eta}}\approx\frac{MU^{2}}{2\eta U^{2}h}=\frac
{M}{2\eta h}, \label{355}%
\end{equation}
and we find $\tau_{\eta}\approx10^{10}%
%TCIMACRO{\unit{s}}%
%BeginExpansion
\operatorname{s}%
%EndExpansion
$. $\tau_{\eta}$ is longer than the designed lifetime of $1$ year, and thus
the relaxation process cannot be established during the experiment.

\textbf{Assembly time}.\textbf{\ }Although we do not discuss assembly of the
film in this paper, we may still consider the restrictions on the assembly
time. To keep the fluid stationary during the assembly precess, the speed of
the enlargement of the film should be much less than the characteristic speed
$1%
%TCIMACRO{\unit{m}}%
%BeginExpansion
\operatorname{m}%
%EndExpansion
/%
%TCIMACRO{\unit{s}}%
%BeginExpansion
\operatorname{s}%
%EndExpansion
$. If we use the velocity $v=0.1%
%TCIMACRO{\unit{m}}%
%BeginExpansion
\operatorname{m}%
%EndExpansion
/%
%TCIMACRO{\unit{s}}%
%BeginExpansion
\operatorname{s}%
%EndExpansion
$, it takes $L/v=10^{4}%
%TCIMACRO{\unit{s}}%
%BeginExpansion
\operatorname{s}%
%EndExpansion
$ to assemble the system. The actual process can be much slower due to
mechanical complexities.

\section{Discussion}

For concreteness we have assumed in our derivation a specific base state of
the system. From the outset we fixed the film thickness, lateral dimensions,
and the expected lifetime. We also chose a particular liquid. However, our
estimates may be used much more generally. We have shown how each of our
quantities depends on the parameters, so that it is a simple matter to infer
the effect of an order-of-magnitude reduction in lateral dimension. This will
reduce the achievable highest Reynolds number according to Eq. (\ref{17}), but
it will reduce the effect of meteoroid impacts, and it is more technically
feasible. Also, larger surface tension and smaller viscosity are advantageous
from the perspective of two-dimensional hydrodynamics. Surface tension and
viscosity are limited by the choice of the liquid, however. In principle, we
can choose the base liquid with the lowest possible viscosity that is allowed
by the acceptable lifetime [Eq. (\ref{2})].

Regarding the film stability against spontaneous rupture, we have based our
estimates entirely on the standard stability analysis. Alternatively, we may
consider a different realization of the stable film configuration, i.e., a
\textquotedblleft black film\textquotedblright\ \cite{4, 32}. Although both
theoretical and experimental studies on such films have been limited to
microscopic bounded films with typical size of a few hundred micrometers
\cite{4}, we think they could be relevant to our space film experiment.
Especially near the boundary where the space film is supported by the external
frame, microscopic theory applies. A sharp transition from an unstable
thinning film to a stable black film may occur at critical film thickness of a
few hundred angstroms. In forming the stable black film, the electrostatic
disjoining pressure due to the repulsive force between the charged surfactant
layers on the two film surfaces contributes to the stability. In addition to
this long range force, the short-range interactions in the adsorption layers
are crucial: the formation of black films depends on the surfactant type and
concentration. The black films are notably much thinner than the films we have
considered under our standard conditions, and thus much less liquid is needed
to form the black films. On the other hand, smaller thickness makes the film
more vulnerable to external impacts and reduces the film lifetime against
evaporation [Eq. (\ref{300})]. Thus we see no clear-cut advantage to the black
films. Moreover, the feasibility of creating a stable black film in our
situation is not clear. In the non-aqueous liquids we are obliged to use, the
prospects for the charged surfaces are much reduced, and thus a stable black
film may not be attainable. Furthermore, there are no assurances that
two-dimensional flows with high enough Reynolds number can be achieved in
black films. Further analysis and investigations of these liquids and their
surfactants are needed in order to address these problems.

A simple film is subject to catastrophic failure when penetrated by
micrometeoroids, and this vulnerability to meteoroids is the chief stumbling
block of our scheme. Thus it is desirable to find a means to arrest the growth
of holes caused by meteoroids. One can imagine designing a liquid film with
this self-healing property. For example, a liquid that immediately solidified
in the vicinity of a small hole would arrest the further opening of the hole.
To achieve this solidification, the liquid would have to contain molecules
that could crosslink the base liquid. The reaction might be triggered by the
energy dissipation of the passing meteoroids. The reaction would have to stop
after solidifying a small region around the hole, so that only a small solid
island remained in the film, and its overall fluidity was little affected.
This scheme appears difficult to achieve but not impossible. Further, recent
experiments \cite{303} showed that millimeter-scale projectiles at impact
speed of $1%
%TCIMACRO{\unit{m}}%
%BeginExpansion
\operatorname{m}%
%EndExpansion
/%
%TCIMACRO{\unit{s}}%
%BeginExpansion
\operatorname{s}%
%EndExpansion
$ may sail through the thin film without leaving a hole due to the pinch-off
effect. These low-velocity results could be relevant in finding effective ways
to protect the space film from micrometeoroids.

Although the potential to achieve a high Reynolds number is one of the most
prominent advantages of the space film experiment, the space film platform
brings other advantages as well. The spatial extent of these films may greatly
exceed the size of terrestrial soap films, even at a size much smaller than
our assumed size of one kilometer. Such films may offer advantages for
studying flow phenomena that require long observation times but not
unprecedented Reynolds numbers. Moreover, the space film can be also useful to
space science itself. Being subject to the external impacts and ambient
forces, these films could be the basic tools to observe and measure such
interplanetary dynamic phenomena as solar wind and cosmic dusts.

Further, although we consider in this paper large-scale thin-film experiment
based on liquids, we can also imagine similar experiments with solid soft
material sheets of large spatial extent in the space environment. Compared to
thin liquid films, such sheets will have less restrictions. The space
environment allows unprecedented size-to-thickness ratio and permits elastic
stress concentration to occur with little disturbance from other forces
\cite{66}. It is important to understand the motions and deformations of large
sheets subject to the ambient conditions in the space environment.

The space film experiment also challenges current space technology. Further
researches and experiments need to be carried out in order to launch and
assemble the system, to observe and measure the system, and to repair and
maintain the system. Similar space programs currently underway suggest that
these logistical challenges are surmountable. For instance, the recent solar
film or space sail experiment that deploys large-scale light solid sheets has
similar properties to our proposed space film, and begins to explore these
techniques \cite{301, 302}. Both the space sail and our proposed film
experiments involve launching and deploying kilometer-scale sheets of light
materials with total weight up to a metric ton, and maintaining the system in
the interplanetary environment for years. We can thus imagine to achieve
significant synergies in technologies and resources among those similar space programs.

\section{Conclusion}

As we have shown, space environment imposes strict conditions on
thermodynamical and other physical properties of large-scale thin liquid
films. Various issues and parameters, such as the film viscosity, vapor
pressure, and compatible surfactants, need to be carefully designed and
experimentally tested to achieve required stability and other desired
properties. We found that despite many potential obstacles to realizing
square-kilometer thin films, only the meteoroid impacts seemingly make large
films unfeasible. Some way, such as a shield or a self-healing mechanism in
film fluid, must be found to prevent meteoroids from destroying the film.
Alternatively we must restrict ourselves to much smaller films, and thus we
must consider possible applications of the space film to the study of
two-dimensional flows of lower Reynolds number.

\begin{acknowledgments}
We would like to acknowledge Elizabeth Goldschmidt for an early study that
helped orient our work. We also thank Roberto Benzi, Francois Blanchette,
Robert Ecke, Sascha Hilgenfeldt, Jean-Marc di Meglio, Toan Nguyen, David
Quere, and William Wheaton for useful discussions. This work was supported in
part by the National Science Foundation's MRSEC Program under Award Number DMR-0213745.
\end{acknowledgments}

\appendix

\section{Possible candidate base liquids}

\textit{Dow Corning }705 (DC 705) pump oil
(penta-phenyl-tri-methyl-tri-siloxane) is a colorless to straw-colored, single
component fluid designed for ultrahigh vacuum applications. Composed of
relatively small molecules in the form of short three-monomer chains ($N=3 $),
it has the highest phenyl content among all DC\ pump oil products and hence
the lowest vapor pressure. Its typical physical and chemical properties are
\cite{29}:%
\[
\begin{tabular}
[c]{|c|c|}\hline
Extrapolated Vapor Pressure, $%
%TCIMACRO{\unit{torr}}%
%BeginExpansion
\operatorname{torr}%
%EndExpansion
$, at $298%
%TCIMACRO{\unit{K}}%
%BeginExpansion
\operatorname{K}%
%EndExpansion
$ & $3\times10^{-10}$\\\hline
Specific Gravity at $298%
%TCIMACRO{\unit{K}}%
%BeginExpansion
\operatorname{K}%
%EndExpansion
$ & $1.09$\\\hline
Viscosity at $298%
%TCIMACRO{\unit{K}}%
%BeginExpansion
\operatorname{K}%
%EndExpansion
$, cSt & $175$\\\hline
Flash Point, open cup, $%
%TCIMACRO{\unit{K}}%
%BeginExpansion
\operatorname{K}%
%EndExpansion
$ & $516$\\\hline
Boiling Point, at $0.5%
%TCIMACRO{\unit{torr}}%
%BeginExpansion
\operatorname{torr}%
%EndExpansion
$, $%
%TCIMACRO{\unit{K}}%
%BeginExpansion
\operatorname{K}%
%EndExpansion
$ & $518$\\\hline
Typical Boiler Temperature $%
%TCIMACRO{\unit{K}}%
%BeginExpansion
\operatorname{K}%
%EndExpansion
$ & $523$ to $543$\\\hline
Surface Tension, $%
%TCIMACRO{\unit{dyn}}%
%BeginExpansion
\operatorname{dyn}%
%EndExpansion
/%
%TCIMACRO{\unit{cm}}%
%BeginExpansion
\operatorname{cm}%
%EndExpansion
$ & $36.5$\\\hline
Heat of Vaporization, $%
%TCIMACRO{\unit{kcal}}%
%BeginExpansion
\operatorname{kcal}%
%EndExpansion
/%
%TCIMACRO{\unit{g}}%
%BeginExpansion
\operatorname{g}%
%EndExpansion%
%TCIMACRO{\unit{mol}}%
%BeginExpansion
\operatorname{mol}%
%EndExpansion
$, at 523$%
%TCIMACRO{\unit{K}}%
%BeginExpansion
\operatorname{K}%
%EndExpansion
$ & $28.2$\\\hline
Molecular Weight & $546$\\\hline
\end{tabular}
\]

Using these data, stability analysis [Eq. (\ref{2})] gives a rupture time of
$100$ years, with proper surfactants added. The typical vapor pressure,
however, is still larger than our earlier estimate [Eq. (\ref{307})] against evaporation.

Vapor pressure can be reduced by lowering temperature. According to the
empirical vapor pressure equation for DC 705 pump oil \cite{29}%
\begin{equation}
\log_{10}P/%
%TCIMACRO{\unit{torr}}%
%BeginExpansion
\operatorname{torr}%
%EndExpansion
=12.31-\frac{6490}{T/%
%TCIMACRO{\unit{K}}%
%BeginExpansion
\operatorname{K}%
%EndExpansion
}\text{,} \label{B1}%
\end{equation}
we find condition Eq. (\ref{307}) can be satisfied at about $273$ $%
%TCIMACRO{\unit{K}}%
%BeginExpansion
\operatorname{K}%
%EndExpansion
.$

To change the temperature will affect the viscosity. Because of their high
phenyl content, pump oils show great change in viscosity with temperature
\cite{30}. In the case of DC 705 pump oil, the viscosity increases from $175$
cSt at $298%
%TCIMACRO{\unit{K}}%
%BeginExpansion
\operatorname{K}%
%EndExpansion
$ to $10000$ cSt at $273%
%TCIMACRO{\unit{K}}%
%BeginExpansion
\operatorname{K}%
%EndExpansion
$. Lowering temperature thus drastically reduces the film fluidity and should
be avoided.

The great sensitivity of viscosity to temperature is an unwelcome property,
and efforts must be made to keep the film temperature stable. Meanwhile, we
can reduce this sensitivity by lowering the phenyl content. For comparison, we
have also considered poly dimethyl-siloxane, or PDMS. No series of organic
liquids show as little change in viscosity with temperature as PDMS \cite{30}.
PDMS are listed as DC 200 fluids, and their viscosity at $298%
%TCIMACRO{\unit{K}}%
%BeginExpansion
\operatorname{K}%
%EndExpansion
$ ranges from $0.65$ cSt to $10^{6\text{ }}$cSt with respect to different
molecular chain lengths. We have particularly considered the PDMS fluid with
viscosity $1000$ cSt, which has a molecular weight of $16500$. Using available
data \cite{31}, we infer that at $273%
%TCIMACRO{\unit{K}}%
%BeginExpansion
\operatorname{K}%
%EndExpansion
$ it has a vapor pressure of $10^{-3}$ $%
%TCIMACRO{\unit{torr}}%
%BeginExpansion
\operatorname{torr}%
%EndExpansion
$, which is far too high compared to the condition Eq. (\ref{307}).

In principle other silicones with low or medium phenyl content can be also
considered as candidate base liquids. They have lower vapor pressure compared
to DC 200 fluids, and their viscosity is not as sensitive to temperature as DC
705 fluid.

We can also reduce the vapor pressure by increasing the number of units $N$.
As shown previously, condition Eq. (\ref{307}) can be satisfied with $N=4$. To
change the molecular weight will also affect the viscosity. According to the
empirical relation \cite{300}:
\begin{equation}
\eta\propto M^{3}, \label{B2}%
\end{equation}
we find for $N=4$ the viscosity increases only modestly to about $450$ cSt,
which is still feasible.

\end{document}